# Autonomic model for self-configuring C#.NET applications

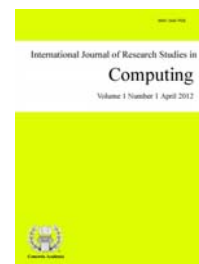

Bassil, Youssef ✉
*Lebanese Association for Computational Sciences, Beirut, Lebanon (youssef.bassil@lacsc.org)*
Semaan, Paul
*Lebanese Association for Computational Sciences, Beirut, Lebanon (paul.semaan@lacsc.org)*

## Abstract

With the advances in computational technologies over the last decade, large organizations have been investing in Information Technology to automate their internal processes to cut costs and efficiently support their business projects. However, this comes to a price. Business requirements always change. Likewise, IT systems constantly evolves as developers make new versions of them, which require endless administrative manual work to customize and configure them, especially if they are being used in different contexts, by different types of users, and for different requirements. Autonomic computing was conceived to provide an answer to these ever-changing requirements. Essentially, autonomic systems are self-configuring, self-healing, self-optimizing, and self-protecting; hence, they can automate all complex IT processes without human intervention. This paper proposes an autonomic model based on Venn diagram and set theory for self-configuring C#.NET applications, namely the self-customization of their GUI, event-handlers, and security permissions. The proposed model does not require altering the source-code of the original application; rather, it uses an XML-based customization file to turn on and off the internal attributes of the application. Experiments conducted on the proposed model, showed a successful automatic customization for C# applications and an effective self-adaption based on dynamic business requirements. As future work, other programming languages such as Java and C++ are to be supported, in addition to other operating systems such as Linux and Mac so as to provide a standard platform-independent autonomic self-configuring model.

# Autonomic model for self-configuring C#.NET applications

## 1. Introduction

Since its inception, Information Technology (IT) has been automating almost every type of manual tasks in everyday life, ranging from industry to education, business, and entertainment. This rapid IT revolution has increased the demand for cutting-edge IT solutions that can contend with every enterprise's business requirements. However, with the ever-growing e-business and the drastic advances and new achievements in computing technologies, IT infrastructures are becoming so complex and heterogeneous to an extent that would make IT people, a few years from now, wrestle to manage, customize, and maintain information systems (Murch, 2004). IBM indicated that the complexity of emerging IT systems is impeding the IT industry from moving to the next millennium of computing; and thus is considered as the most significant forthcoming challenge facing the world of information technology (Horn, 2001). In fact, IT systems have already started causing both time and money lost as IT administrators are always fighting to customize and configure software applications that constantly require updates and modifications, particularly when they are being operated by different types of users, in different environments, for different business requirements, and under different functional constraints. Autonomic computing can cope with this challenge as they are self-configuring, self-healing, self-optimizing, and self-protecting (Murch, 2004). In effect, autonomic systems are meant to automate laborious IT processes and make them autonomous enough to self-adapt according to the mutable business requirements and to the dynamic changes in computing environments.

This paper proposes a new autonomic model for self-configuring C#.NET Windows computer applications. It is mathematically based on Venn diagram and set theory, and it predominantly supports the self-customization of three elements: graphical user interfaces (GUI), event-handlers mainly representing program's functionalities, and security permissions. Practically, self-customization is achieved by deploying an application with a customization file written in XML language that contains customization specifications. The purpose of the proposed model is to automate the customization of C# applications with the least manual intervention and to discharge IT people from the load of building different versions for the same application. As a result, this would increase company's profitability and boost the working capacity of the administrative staff.

## 2. Autonomic System

An autonomic computing system is a proactive system that automatically adjusts and adapts itself according to its computing environment, relieving IT administrators from performing laborious and manual maintenance tasks (Parashar & Hariri, 2007). Essentially, an autonomic computing system has four properties: self-configuring, self-healing, self-optimizing, and self-protecting (Murch, 2004; Parashar & Hariri, 2007).

➢ Self-Configuring: An autonomic system is able to automatically modify its configuration according to the settings of the computing environment, such as automatically downloading and installing software patches and service packs, updating virus signatures, and setup software automatically.

➢ Self-Healing: An autonomic system is able to automatically detect errors at runtime and recover from failures without any human intervention. Self-healing can be achieved by using self-restarting and fault-tolerant components, and creating backup plans for prospective failure scenarios.

➢ Self-Optimizing: An autonomic system is able to automatically tune and adapt itself and its resources based on the state of its execution environment. It includes increasing automatically disk storage to handle extra data, and allocating extra processing cycles for computationally demanding applications.

➢ Self-Protecting: An autonomic system is able to automatically detect and defend against security attacks, threats, and unauthorized system access. It includes using encryption algorithms to secure data transmission, installing intrusion detection systems and firewalls, and enforcing system policies and user access controls.

## 3. Motivations for Self-Customization

GUI short for Graphical User Interface is a visual program interface composed of graphical objects and elements such as windows, menus, buttons, textboxes, and icons, organized into a certain layout to allow computer users to easily interact and manage the underlying functionalities and features of a particular application. GUI interfaces are currently available on many devices such as personal computers, PDAs, smart-phones, electronic home appliances, and proprietary information systems such as retail store systems, payroll systems, and banking systems (Tidwell, 2010).

On the other hand, an event in a computing context is an action that is usually triggered outside the scope of a program and handled by a piece of code inside the program itself, known as event-handler. Basically, an event-handler is a procedure or function containing a piece of code that is executed to handle and act in response to the application's external events such as mouse clicks and keyboard keystrokes. In effect, event-driven systems promote interactivity and ease the communication between the man and the machine, a concept often referred to as Man-Machine Interface (Scott, 2009).

In a different context, security permission is an authorization to perform an operation, such as writing, on a specific object, such as file. Permissions can be granted to or revoked from computer users and applications, and they control and dictate which user can do, what a user can do, and what an application can have access to (Bishop, 2004). As information systems are being used by different classes of users each with different positions, roles, and privileges, and belonging to different departments, there is a necessity to find a way to automatically customize their GUI, event-handlers, and security permissions so as to make them compatible and suitable for every user. For instance, in a web hosting application, customers can purchase domains through an online web-based GUI form. However, they cannot cancel an already purchased domain and get refunded unless they contact the support team. This backend support team is provided with a different interface connected to different functionalities and having different security permissions that allow him to perform administrative tasks. Furthermore, customers who are asking for a discount or a promotional deal should refer to the sales department which has access to another GUI interface whose prime functionality is to unlock the original price and issue discounts.

As a result, IT people and software engineers are obliged to manually build different versions for the same application so as to keep pace with this degree of mix constraints and requirements. It is then crucial to automate the customization process of computer applications and allow them to dynamically self-adapt their GUI, functionalities, and security permissions in accordance with the always changing business rules, requirements, and security policies.

## 4. Related Work

Several solutions were proposed to autonomously self-customize user interfaces, some of them are based on dynamic self-configuration; while others are based on transformation models. For instance, Penner (2009) proposed a tool that integrates various subsystems and automatically generates and configures graphical user interfaces. The approach employs object-oriented models of interface elements and interactions, managed by an algorithm that dynamically generates user interfaces pertaining to particular users. Scholaert (2010) proposed an autonomous man-machine interface for a communication device in a network environment that can be dynamically updated according to predefined computing conditions. The man-machine interface is constantly updated to reflect services offered by service providers in a specific operational environment. Pohjalainen (2010) proposed a method for self-configuration of user interface via software introspection combined with semantic mapping of backend methods that ease the management of user interface during the development process. Lift (Chen-Becker, Weir, & Danciu, 2009) is a web programming framework that employs a linking engine that uses templates to transform scripting instructions into web pages. Besides, Lift offers several advanced utilities and functions to automatically customize the output being generated.

Kuikka and Penttonen (1993) proposed a transformation method based on document type definition (DTD) which converts document instances originally bound to a generic DTD into another type of documents. In this approach an already existing document X along with its grammar Y and a new grammar Y' are fed to the system. The system then converts document X into a corresponding document X' that is compatible with the grammar Y'. In this way, a web graphical interface can be dynamically generated from a DTD using the appropriate grammar Y' for web documents. Amaneddine, Bahsoun, and Bodeveix (2004) proposed the TransM system which is a structured document transformation model whose purpose is to convert an instance document conforming to a DTD into another instance conforming to a another DTD derived from the transformation rules of the TransM system. Additionally, the system harnesses the XSLT transformation language to format and apply styles on the output document that eventually can be rendered as a web interface. Bird, Goodchild, and Halpin (2000) proposed an approach to transform meta-models into XML Schemas and vice versa with minimum data redundancy. The input model is versatile in that it may describe any type of systems including GUI web interfaces. For this purpose, an algorithm based on twelve heuristic rules is used to assist in the conversion of metadata into an XML Schema and consequently into a graphical web interface.

## 5. Proposed Solution

This paper proposes a new autonomic model for self-configuring C#.NET Windows applications. It consists of self-customizing GUI interfaces, event-handlers representing program's functionalities, and security permissions of C#.NET programs using a proprietary XML customization language. Practically, every application to customize is to be deployed alongside with a customization file written in XML language that contains customization specifications to automatically configure the application at runtime. For integrity purposes, the XML customization file is always validated against a DTD (Document Type Definition) file to ensure whether or not it conforms to the original specifications of the XML customization language. In essence, XML short for Extensible Markup Language is an open standard plaintext portable language for exchanging data between computing systems, made out of elements and attributes that define the semantics of data. XML is case-sensitive and is understandable by both humans using a traditional text editor and computers using a special XML parser (Deitel, 2007). The novelty of the proposed model is that it does not require altering the source-code of the original application; rather, it just involves adding a tiny external generic proxy module to the application so that it handles all the complex underlying customization logic.

### 5.1 Self-Customization of GUI

The proposed model supports the customization of several C# GUI controls that are originally part of the .NET Framework and found in the "System.Windows.Forms" namespace. These controls are respectively: "Label", "Textbox", "Button", "ToolStripMenuItem", "Form", "Checkbox", "RadioButton", "ComboBox", and "PictureBox". Principally, these controls can be customized through their "Visible", "Enabled", and "Text" properties. Besides, the "Checkbox" control can be customized through its "Checked" property, the "PictureBox" through its "Image" property, and the "ComboBox" through its "Items" property. Figure 1 depicts all the C# user interface controls along their properties that can be customized dynamically using the proposed model.

The XML language for GUI customization consists of W3C standard XML markups that correspond to the actual customization instructions. It is defined as follows:

```
<customization>
   <GUI>
      <control>
         <name>Control Name</name>
         <type>Control Type</type>
         <property>Property to Customize</property>
         <value>Customizing Value</value>
      </control>
   </GUI>
</customization>
```

The XML customization language starts with the "customization" tag which indicates that the file is meant for customization purposes. It actually contains a child tag, namely the "GUI" tag for specifying user interface customizations. The "GUI" tag has one or more children tags called "control" that each specifies a particular user interface control to be customized. Each "control" tag defines four children tags: the "name" tag designating the name of the control to customize, the "type" tag designating the type of the control to customize, the "property" tag designating the property through which the control is to be customized, and the "value" tag designating the customizing value to be assigned to the property of the control to customize.

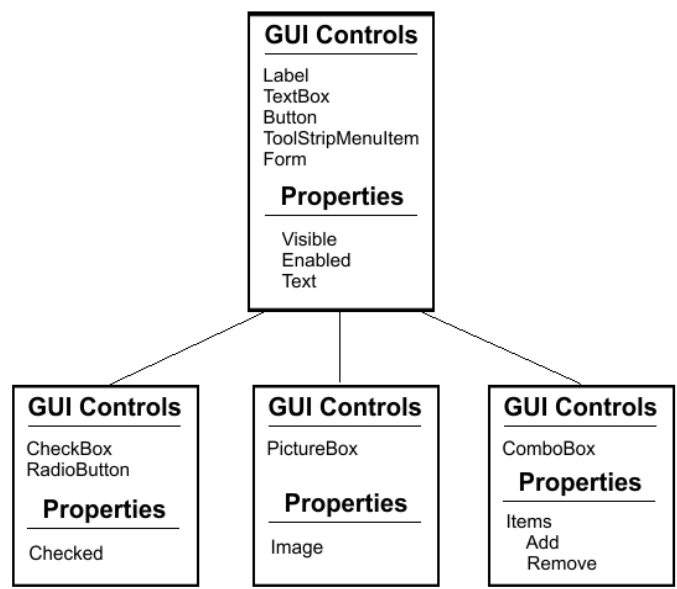

***Figure 1***. Customizable C# GUI Controls

*5.2 Self-Customization of Event-Handlers*

The proposed model supports the customization of C# event-handlers only pertaining to the following C# GUI controls: "Label", "Textbox", "Button", "ToolStripMenuItem", "Form", "Checkbox", "RadioButton", "PictureBox", and "ComboBox". Basically, every event-handler represents a function or procedure that can be enabled or disabled through the XML customization language. Figure 2 depicts the different C# GUI controls whose event-handlers can be customized dynamically using the proposed model.

The XML language for event-handlers customization consists of W3C standard XML markups that correspond to the actual customization instructions. It is defined as follows:

```
<customization>
     <EVENTS>
          <event>
                <name>Name of the Handler to Customize</name>
                <controlName>Corresponding GUI Control</controlName>
                <action>+ | −</action>
          </event>
     </EVENTS>
</customization >
```

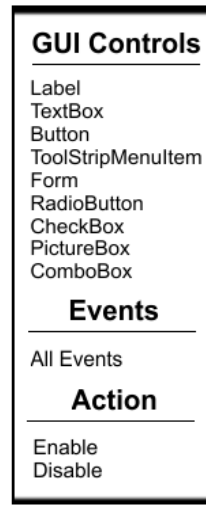

**Figure 2**. Customizable Event-Handlers

The XML language for event-handlers customization starts with a "customization" tag indicating that this file is meant for customization purposes. It actually contains a child tag, namely the "EVENTS" tag for specifying event-handlers customizations. The "EVENTS" tag has one or more children tags called "event" that each specifies a particular event-handler to be customized. Every "event" tag defines three children tags: the "name" tag designating the name of the event-handler to customize, the "controlName" tag designating the corresponding GUI control bound to the event-handler to customize, and the "action" tag designating whether the event-handler should be enabled or disabled. A plus sign (+) means enable, whereas a minus sign (-) means disable.

*5.3  Self-Customization of Security Permissions*

The proposed model supports the customization of three different C# security permissions: "Disk Access", "Network Access", and "Process Access". Mainly, permissions can only be granted to or revoked from a GUI "Form" control. The "Form" control is the largest container in a C# application which houses all other GUI controls and the code-behind of all event-handlers, as well as other functions and variables. Consequently, changing its security permissions will result in changing the security access control for the whole program. Figure 3 depicts the three security permissions that can be customized dynamically using the proposed model.

The XML language for security permissions customization consists of W3C standard XML markups that correspond to the actual customization instructions. It is defined as follows:

```
<customization>
     <PERMISSIONS>
          <permission>
                <name>Permission Name</name>
                <action>+ | −</action>
          </permission>
     </PERMISSIONS>
</customization>
```

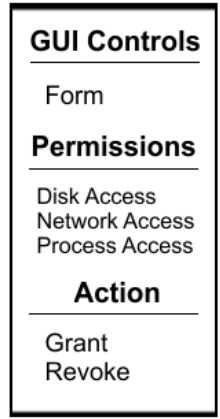

*Figure 3.* Customizable Security Permissions

The XML language for security permissions customization starts with a "customization" tag that indicates that this file is meant for customization purposes. It actually contains a child tag, namely the "PERMISSIONS" tag for specifying permissions customizations. The "PERMISSIONS" tag has one or more children tags called "permission" that each specifies a particular permission to be customized. Every "permission" tag defines two children tags: the "name" tag designating the name of the permission to customize and the "action" tag designating whether the permission should be granted or revoked. A plus sign (+) means grant, whereas a minus sign (-) means revoke.

*5.4 DTD Validation*

In order to validate the XML language of the customization file deployed with the application to customize, a Document Type Definition (DTD) is used for this purpose. The DTD is a W3C standard markup document containing declarations that certify that the XML customization file is compliant with the specifications of the customization language. The DTD document is defined as follows:

```
<!ELEMENT customization (GUI, EVENTS, PERMISSIONS)>
<!ELEMENT GUI (control*)>
<!ELEMENT control (name, type, property, value)>
<!ELEMENT name (#PCDATA)>
<!ELEMENT type (#PCDATA)>
<!ELEMENT property (#PCDATA)>
<!ELEMENT value (#PCDATA)>
<!ELEMENT EVENTS (event*)>
<!ELEMENT event (name, controlName, action)>
<!ELEMENT name (#PCDATA)>
<!ELEMENT controlName (#PCDATA)>
<!ELEMENT action (+|-)>
<!ELEMENT PERMISSIONS (permission*)>
<!ELEMENT permission (name, action)>
<!ELEMENT name (#PCDATA)>
<!ELEMENT action (+|-)>
```

## 6. The Mathematical Model

The proposed model is at heart based on Venn diagram (Venn, 1880) and set theory (Georg & Jourdain, 1955) of discrete mathematics. By definition, a Venn diagram is a representation of all hypothetically possible logical relations between finite collections of sets. It was originally used as an application for set theory to calculate intersection, union, complement, and Cartesian product between different sets. Mathematically, the proposed model can be represented as having three different sets: The application to be customized denoted by "*Application to Customize*", the XML customization file containing customization instructions denoted by "*XML Customization File*", and the output application after being customized denoted by "*Customized Application*". Every one of these sets is a union of three types of elements, denoted respectively by "*GUI-Elements*", "*Events-Elements*", and "*Permissions-Elements*". Formally, the proposed model is denoted by M and is represented as follows:

> *M= ("Application to Customize", "XML Customization File", "Customized Application")*

where:

> *Application to Customize = { GUI-Elements U Events-Elements U Permissions-Elements }*
> *XML Customization File = { GUI-Elements' U Events-Elements' U Permissions-Elements' }*
> *Customized Application = { GUI-Elements" U Events-Elements" U Permissions-Elements" }*

A Venn diagram representation for the model M is depicted in Figure 4.

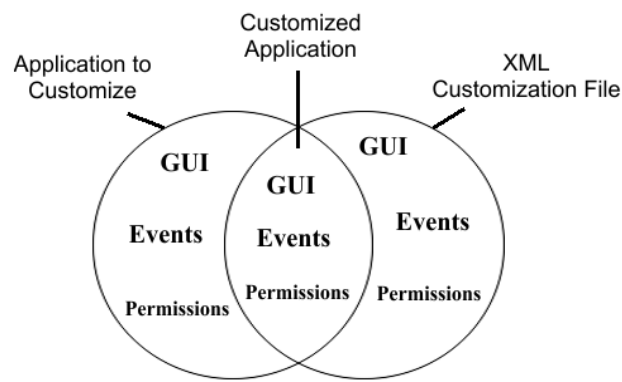

***Figure 4.*** The Proposed Model as Venn diagram

Accordingly, two fundamental properties can be drawn from the above diagram: the first one is Intersection; while the second is Subset. Obviously, the intersection between the "*Application to Customize*" and the "*XML Customization File*" sets is equal to the "*Customized Application*" set. Subsequently, the "*XML Customization File*" is a subset of the "*Application to Customize*". These two properties are valid as the elements to customize in the XML customization file are originally part of the application to customize. Hence, the following two properties stand:

> *1. Customized Application = Application to Customize ∩ XML Customization File*
> *where: { x: x Є Application to Customize AND x Є XML Customization File }*

> *2. XML Customization File ⊆ Application to Customize*
> *where: { x: x Є XML Customization File IS IN Application to Customize }*

For illustration purposes, let's assume the following:

> *Application to Customize = (GUI = {a, b, c}) U (Events = {m, n, p}) U (Permissions = {s, t, v})*
> *XML Customization File = (GUI = {b}) U (Events = {n, p}) U (Permissions = {v})*

The proposed model first performs intersection between the "*Application to Customize*" and the "*XML Customization File*" sets, and then it performs union on the intersection results. The output is the "*Customized Application*" which can be expressed as:

> *Customized Application = ({a, b, c} ∩ {b}) U ({m, n, p} ∩ {n, p}) U ({s, t, v} ∩ {v})*
> *Customized Application = {b} U {n, p} U {v}*
> *Customized Application = (GUI = {b}) U (Events = {n, p}) U (Permissions = {v})*

Figure 5 shows the big image of the complete mathematical model expressed in set theory.

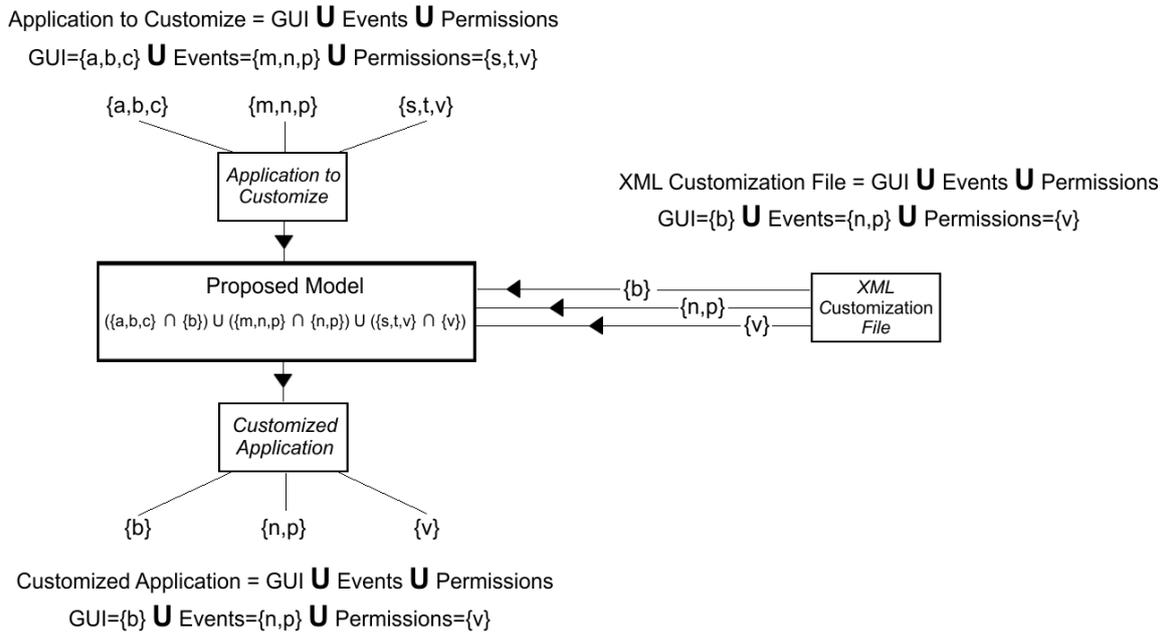

**Figure 5.** The Complete Mathematical Model

## 7. Experiments and Results

In the experiments, the proposed model was evaluated; a simple computer application was built using C#.NET and deployed on a Windows PC with Internet access. By default, this sample application contains several GUI controls, an advertisement image in the footer section of the program, and three security permissions. Formally, it comprises three entities: the original application which is meant to be customized denoted by "*Application to Customize*", the XML file containing the customization instructions denoted by "*XML Customization File*", and the final output customized application denoted by "*Customized Application*".

### 7.1 The Application to Customize

The "*Application to Customize*" is a simple application for creating a new user account in a stock management system. Majorly, the application is composed of GUI controls, event-handlers, and security permissions, and hence it can be formally expressed as:

*Application to Customize = **GUI** U **Events** U **Permissions***

where:

> ***GUI*** = { fileToolStripMenuItem, adminToolStripMenuItem, firstnameLabel, firstnameTextbox, lastnameLabel, lastnameTextbox, countryLabel, countryCBX, priceLabel, priceTextbox, genderLabel, maleRadio, femaleRadio, approveCheckbox, createButton, saveButton, advertisementImage, mainForm }
> ***Events*** = { createButton_Click, saveButton_Click, fileToolStripMenuItem_Click, admin ToolStripMenuItem_Click }
> ***Permissions*** = { DiskAccess, NetworkAccess, ProcessAccess }

Figure 6 shows the main GUI interface of the application to customize.

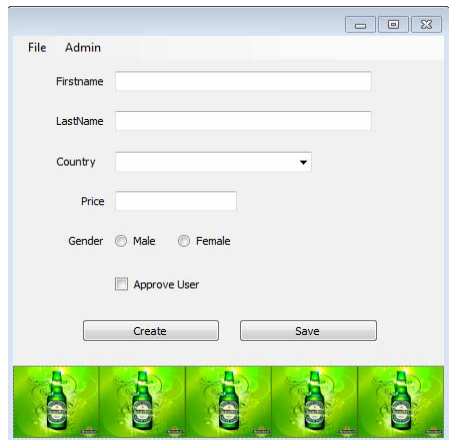

**Figure 6.** Application to Customize

*7.2 The XML Customization File*

The "*Application to Customize*" is designed to be operated by two types of users: regular users and administrators, each having different GUI, event-handlers, and security permissions. For this reason, two XML customization files are required: one deployed on the user machine called "*user.xml*" and one deployed on the administrator machine called "*admin.xml*". Both files can be formally expressed as:

> *XML Customization File = **GUI** U **Events** U **Permissions***

where the "*user.xml*" comprises the following sets:

> **GUI** = { *fileToolStripMenuItem, firstnameLabel, firstnameTextbox, lastnameLabel, lastnameTextbox, countryLabel, countryCBX, priceLabel, priceTextbox, genderLabel, maleRadio, femaleRadio, createButton, advertisementImage, mainForm* }
> **Events** = { *createButton_Click, fileToolStripMenuItem_Click* }
> **Policies** = { }

and the "*admin.xml*" comprises the following sets:

> **GUI** = { *adminToolStripMenuItem, firstnameLabel, firstnameTextbox, lastnameLabel, lastnameTextbox, countryLabel, countryCBX, priceLabel, priceTextbox, genderLabel, maleRadio, femaleRadio, approveCheckbox, saveButton, mainForm* }
> **Events** = { *saveButton_Click, admin ToolStripMenuItem_Click* }
> **Policies** = { *DiskAccess, NetworkAccess* }

Implementation-wise, the "*user.xml*" customization file is defined as follows:

```
<customization>
   <GUI>
      <control>
         <name>adminToolStripMenuItem</name>
         <type>ToolStripMenuItem</type>
         <property>Visible</property>
         <value>False</value>
      </control>
      <control>
         <name>approvedCheckbox</name>
         <type>CheckBox</type>
         <property>Visible</property>
         <value>False</value>
      </control>
```

```
        <control>
            <name>saveButton</name>
            <type>Button</type>
            <property>Visible</property>
            <value>False</value>
        </control>
    </GUI>

    <EVENTS>
        <event>
            <eventName>saveButton_Click</eventName>
            <controlName>saveButton</controlName>
            <action>-</action>
        </event>
        <event>
            <eventName>adminToolStripMenuItem_Click</eventName>
            <controlName>adminToolStripMenuItem</controlName>
            <action>-</action>
        </event>
    </EVENTS>

    <POLICIES>
        <policy>
            <name>DiskAccess</name>
            <action>-</action>
        </policy>
        <policy>
            <name>NetworkAccess</name>
            <action>-</action>
        </policy>
        <policy>
            <name>ProcessAccess</name>
            <action>-</action>
        </policy>
    </POLICIES>
</customization>
```

while the "*admin.xml*" customization file is defined as follows:

```
<customization>
    <GUI>
        <control>
            <name>fileToolStripMenuItem</name>
            <type>ToolStripMenuItem</type>
            <property>Visible</property>
            <value>False</value>
        </control>
        <control>
            <name>createButton</name>
            <type>Button</type>
            <property>Visible</property>
            <value>False</value>
        </control>
        <control>
            <name>advertisementImage</name>
            <type>Image</type>
            <property>Visible</property>
            <value>False</value>
        </control>
    </GUI>
```

```
<EVENTS>
    <event>
        <eventName>createButton_Click</eventName>
        <controlName>createButton</controlName>
        <action>-</action>
    </event>
    <event>
        <eventName>fileToolStripMenuItem_Click</eventName>
        <controlName>fileToolStripMenuItem</controlName>
        <action>-</action>
    </event>
</EVENTS>

<POLICIES>
    <policy>
        <name>ProcessAccess</name>
        <action>-</action>
    </policy>
</POLICIES>
</customization>
```

### 7.3 The Customized Application

In order to customize each of the applications, intersection is performed between the "*Application to Customize*" and the "*XML Customization File*" sets. This operation can be formally expressed as:

$$\textit{Customized Application} = \textbf{\textit{Application to Customize}} \cap \textbf{\textit{XML Customization File}}$$

The results are two different applications each with different GUI, event-handlers, and security permissions. They can both be formally expressed as:

$$\textit{Customized Application} = \textbf{\textit{GUI U Events U Permissions}}$$

The "*User*" application now comprises the following sets:

> **GUI** = { fileToolStripMenuItem, firstnameLabel, firstnameTextbox, lastnameLabel, lastnameTextbox, countryLabel, countryCBX, priceLabel, priceTextbox, genderLabel, maleRadio, femaleRadio, createButton, advertisementImage, mainForm }
> **Events** = { createButton_Click, fileToolStripMenuItem_Click }
> **Policies** = { }

while the "*Admin*" application comprises the following sets:

> **GUI** = { adminToolStripMenuItem, firstnameLabel, firstnameTextbox, lastnameLabel, lastnameTextbox, countryLabel, countryCBX, priceLabel, priceTextbox, genderLabel, maleRadio, femaleRadio, approveCheckbox, saveButton, mainForm }
> **Events** = { saveButton_Click, admin ToolStripMenuItem_Click }
> **Policies** = { DiskAccess, NetworkAccess }

Figure 7 shows the user "*Customized Application*", while figure 8 shows the admin "*Customized Application*".

## 8. Conclusions and Future Work

This paper presented an autonomic model based on Venn diagram and set theory for self-configuring C#.NET Windows applications. The model supports self-customization of GUI interfaces, event-handlers, and security permissions through a customization file specified using the XML language. First, the autonomic customization of GUI interfaces implies that an application is no more bound to a specific static GUI interface and layout; rather, it is adaptable dynamically according to the application's requirements. Second, the autonomic customization of event-handlers implies that an application is no more compiled with a predefined set of functions and procedures; rather, its internal functionalities can be turned on and off based on users' tasks. Third, the autonomic customization of security permissions implies that an application is no more designed to operate by a particular group of users; rather, the same application can be deployed on dissimilar machines having dissimilar level of security, permissions, and privileges. As a result, IT people are relieved from the burden of constantly maintaining, managing, and developing customized applications to meet the ever-changing business requirements. Consequently, the autonomic proposed model can reduce the overall development and maintenance cost and time, and simplify the implementation of rapidly growing and complex computing infrastructures.

As future work, more investigations are to be carried out on other programming languages such as Java and C++, in addition to other operating systems such as Linux and Mac, in an attempt to deliver a platform-independent autonomic self-configuring model. Furthermore, and since web applications are becoming common these days, methods are to be researched to provide self-customization for web elements such as web forms, web GUI, and client-side scripts in web information systems.

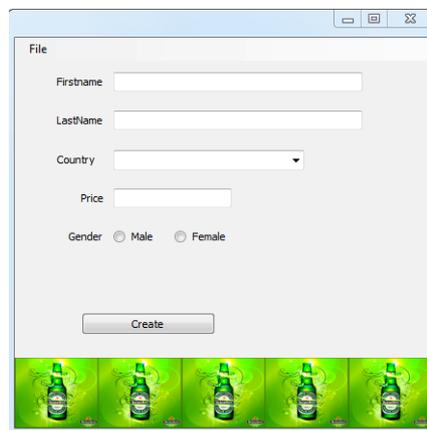

*Figure 7.* User Customized Application

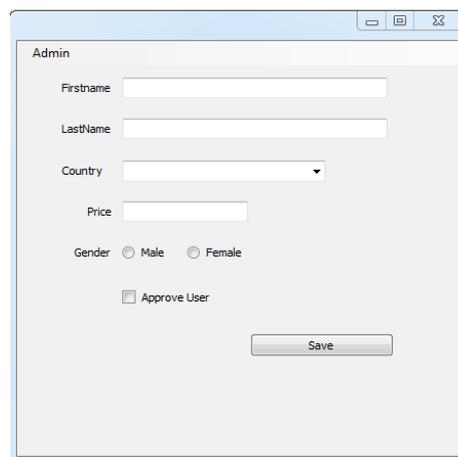

*Figure 8.* Admin Customized Application

***Acknowledgement:*** This research was funded by the Lebanese Association for Computational Sciences (LACSC), Beirut, Lebanon under the "Autonomic Computing Research Project – ACRP2011".

## 9. References:

Amaneddine, N., Bahsoun, J. P., & Bodeveix, J. P. (2004). *TransM: A structured document transformation model.* In the 3rd International Conference Information Systems Technology and its Applications, Salt Lake City, Utah.

Bird, L., Goodchild, A., & Halpin, T. (2000). *Object role modeling and XML-schema*. In International Conference on Conceptual Modeling, Salt Lake City, UT.

Bishop, M. (2004). *Introduction to computer security*. Reading, MA: Addison-Wesley Professional.

Chen-Becker, D., Weir, T., & Danciu, M. (2009). *The definitive guide to lift: A scala-based web framework*. Berkeley, CA: Apress.

Deitel, P. (2007). *Internet & World Wide Web: How to program* (4th ed.). Upper Saddle River, NJ: Prentice Hall.

Georg, C., & Jourdain, P. (1955). *Contributions to the founding of the theory of transfinite numbers*. New York: Dover.

Horn, P. (2001). *Autonomic computing, IBM's perspective on the state of information technology.* IBM Corporation.

Kuikka, E., & Penttonen, M. (1993). Transformation of structured documents with the use of grammar. *Electronic Publishing, 6*(4), 373-383.

Murch, R. (2004). *Autonomic computing*. Englewoods cliffs, NJ: IBM Press and Prentice Hall.

Parashar, M., & Hariri, S. (Eds.). (2007). *Autonomic computing: Concepts, infrastructure, and applications.* Boca Ranton, FL: CRC Press.

Penner, R. (2009). *Self-configuring user interface design.* Minneapolis, MN: Iterativity, Inc.

Pohjalainen, P. (2010). *Self-configuring user interface components.* In the First Workshop on Semantic Models for Adaptive Interactive Systems.

Scholaert, H. (2010). *Self-configuring man-machine interface for a communication terminal*. United States Patent No: 20100248719.

Scott, M. (2009). *Programming language pragmatics* (3rd ed.). San Francisco, CA: Morgan Kaufmann.

Tidwell, J. (2010). *Designing interfaces* (2nd ed.). Cambridge, MA: O'Reilly Media.

Venn, J. (1880). On the diagrammatic and mechanical representation of propositions and reasonings. *Dublin Philosophical Magazine and Journal of Science, 10*(50), 1-18.